\documentstyle[11pt]{article}
\newlength{\bredde}
\def\slash#1{\settowidth{\bredde}{$#1$}\ifmmode\,\raisebox{.15ex}{/}
\hspace*{-\bredde} #1\else$\,\raisebox{.15ex}{/}\hspace*{-\bredde} #1$\fi}
\newcommand{\beq}{\begin{equation}}
\newcommand{\eeq}{\end{equation}}
\newcommand{\noi}{\vspace{12pt}\noindent}
\newcommand{\lG}{\raise.3ex\hbox{$\stackrel{\leftarrow}{G}$}}
\newcommand{\lU}{\raise.3ex\hbox{$\stackrel{\leftarrow}{U}$}}
\newcommand{\lP}{\raise.3ex\hbox{$\stackrel{\leftarrow}{{\cal P}}$}}
\newcommand{\leta}{\raise.3ex\hbox{$\stackrel{\leftarrow}{\eta}$}}
\newcommand{\lOmega}{\raise.3ex\hbox{$\stackrel{\leftarrow}{\Omega}$}}
\newcommand{\ldr}{\raise.3ex\hbox{$\stackrel{\leftarrow}{\delta^r}$}}

\def\beqn{\begin{eqnarray}}
\def\eeqn{\end{eqnarray}}

\def\gtwid{\raise.3ex\hbox{$>$\kern-.75em\lower1ex\hbox{$\sim$}}}
\def\ltwid{\raise.3ex\hbox{$<$\kern-.75em\lower1ex\hbox{$\sim$}}}

\def\la{\lambda}
\def\tila{\tilde{\lambda}}
\begin{document}
\topmargin -1.4cm
\oddsidemargin -0.8cm
\evensidemargin -0.8cm
\title{\Large{{\bf QCD Dirac Spectra With and Without Random Matrix Theory}}}
\thanks{
Invited talk presented at the Fourth Workshop on Quantum Chromodynamics,
Paris, June 1-6, 1998. To appear in those proceedings.} 
\vspace{1.5cm}
\author{~\\~\\
{\sc Poul H. Damgaard}\\
The Niels Bohr Institute\\ Blegdamsvej 17\\ DK-2100 Copenhagen\\
Denmark}
\maketitle
\vfill
\begin{abstract}
Recent work on the spectrum of the Euclidean Dirac operator spectrum show 
that the exact microscopic spectral density can be computed in both random 
matrix theory, and directly from field theory. Exact relations to effective 
Lagrangians with additional quark species form the bridge between the two 
formulations. Taken together with explicit computations in the chGUE random 
matrix ensemble, a series of universality theorems are used to prove 
that the finite-volume QCD partition function coincides exactly with the 
universal double-microscopic limit of chUE random matrix partition functions. 
In the limit where $N_f$ and $N_c$ both go to infinity with the ratio 
$N_f/N_c$ fixed, the relevant effective Lagrangian undergoes a third order 
phase transition of Gross-Witten type. 
\end{abstract}
\vfill
\begin{flushleft}
NBI-HE-98-17 \\
hep-th/9807026
\end{flushleft}
\thispagestyle{empty}
\newpage

\setcounter{page}{1}
\section{Introduction}

Over the last five years it has gradually become clear that the eigenvalue 
spectrum of the Dirac operator in Euclidean QCD (and other gauge theories) 
can be
computed {\em exactly} in a particular finite-volume scaling region. 
The origin of these developments dates back to work in the 1980's on QCD 
in a finite volume (see $e.g.$ ref. \cite{GL}), but the main breakthrough 
came with two very influential papers in 1992 by Leutwyler and Smilga 
\cite{LS}, and by Shuryak and Verbaarschot \cite{SV}. In the latter paper 
an intriguing relation to random matrix theory was pointed out for the first 
time, and this led quickly to a series of theoretical developments that 
clarified the connection between Dirac eigenvalue spectra in gauge theories and
random matrix theory \cite{V,VZ,ADMN} (see also the reviews 
of ref. \cite{Vrev}). Here we shall discuss some very recent developments.

\noi 
The central object of study is the spectral density 
$\rho(\lambda;m_1,\ldots,m_{N_{f}}) =
\sum_n \langle\delta(\lambda - \lambda_n)\rangle_{\nu}$, where the average
is taken over all gluon configurations with fixed topological charge $\nu$,
and where the Dirac eigenvalues $\lambda_n$ are solutions to $\slash{D}
\phi_n = \lambda_n\phi_n$. In a finite volume $V$ it is convenient to
introduce instead a rescaled, {\em double-microscopic}, spectral 
density \cite{SV}
\beq
\rho_S(\zeta;\mu_1,\ldots,\mu_{N_{f}}) 
~\equiv~ \frac{1}{V\Sigma}\rho\left(\frac{\zeta}{V\Sigma};
\frac{\mu_1}{V\Sigma},\ldots,\frac{\mu_{N_{f}}}{V\Sigma}\right)
~,~~~~~~~~ V ~\to~ \infty ~,
\eeq
which computes the local density near $\lambda \sim 0$, on a scale set by
the chiral condensate $\Sigma$, as computed in the massless theory. In the 
infinite-volume limit $\Sigma$ will be proportional to $\rho(0)$, but
the microscopic spectral density will generically vanish at $\zeta=0$,
in accordance with the fact that in any finite volume $V$ there is no
spontaneous chiral symmetry breaking. 

\noi
Consider now a finite-volume range 
$1/\Lambda_{\mbox{\rm QCD}} << L << 1/m_{\pi}$, where $L \sim V^{1/4}$.
The essential observation of 
Leutwyler and Smilga \cite{LS} was that in a sector of fixed topological
charge $\nu$ this actually defines a finite-size {\em scaling region}.
What this means becomes clear when one considers the finite-volume
partition function (really the generating function for the chiral 
condensate). In the above limit it equals
\beq
{\cal Z}_{\nu}^{(N_{f})}(\mu_1,\ldots,\mu_{N_{f}}) ~=~ 
\int dU (\det U)^{\nu}\exp
\left[V\Sigma~ {\mbox{\rm Re Tr}} [{\cal M}U]\right] \label{ZLS}
\eeq
where the integral is taken over $U(N_f)$, and ${\cal M}$ is the quark
mass matrix, which we take to be diagonal in the masses $m_i$. 
This generating function depends only on one very particular
combination, $\mu_i \equiv m_i V\Sigma$, and is in this sense a {\em
scaling function}. The only needed ingredient is the existence of a 
non-vanishing chiral condensate $\Sigma$. Although the effective Lagrangian
is consistent with the Gell-Mann--Oakes--Renner relation, the partition
function does not depend on other dimensionful parameters, such as $f_{\pi}$. 
Of course, the above representation becomes exact only in the limit, 
but this is precisely 
what we mean by having an exact finite-size scaling function: the universal 
result can be recovered to any required accuracy by tuning $V$. 
Corrections are suppressed by powers of $1/V$; we will comment on the 
nature of such corrections below. 

\noi
The finite-volume partition function (\ref{ZLS}) has been evaluated
exactly for an arbitrary mass matrix ${\cal M}$ and for any $N_f$ and $\nu$ 
\cite{JSV}. It turns out that complete spectral information about the Dirac
operator in the double-microscopic limit can be obtained from this
effective partition function alone \cite{DAD,OTV}. Curiously, the simplest
starting point is the random matrix theory formulation of the finite-volume
partition function (\ref{ZLS}) \cite{SV}:
\beq
\tilde{\cal Z}_{\nu}^{(N_{f})}(m_1,\ldots,m_{N_{f}}) 
~=~ \int\! dW \prod_{f=1}^{N_{f}}{\det}\left(M + m_f\right)~
\exp\left[-\frac{N}{2} {\rm tr}\, V(M^2)\right] ~,
\label{ZRM}
\eeq
where
\beq
M ~=~ \left( \begin{array}{cc}
              0 & W^{\dagger} \\
              W & 0
              \end{array}
      \right) ~.
\eeq
Here $W$ is a rectangular complex matrix of size
$N\times(N\! +\! |\nu|)$. In the large-$N$ limit the space-time volume $V$ 
of QCD is identified with $2N$. The potential $V(M^2)$ can be parametrized
in a general way by $V(M^2) = \sum (g_k/k)M^{2k}$. It was proven in
refs. \cite{ADMN,DN} that all double-microscopic spectral correlators
(including, as the most simple case, the double-microscopic spectral density
itself) in fact are {\em universal}, i.e. independent of the choice of
$V(M^2)$ up to a rescaling of the local macroscopic spectral density, 
here $\rho(0)$. The class of potentials that fall into this specific
universality class is huge, its boundary given by potentials
for which one has $\rho(0) = 0$, but $\rho^{(2n)}(0) \neq 0$
for some integer $n$ \cite{ADMN'}.

\noi
To begin, the random matrix theory computations were always performed in
the specific case of a Gaussian potential. Rewriting that particular
random matrix theory partition it was shown already in refs. \cite{SV,HV}
that in the double-microscopic limit in which $\zeta \equiv
\lambda N2\pi\rho(0)$ and $\mu_i \equiv m_i N2\pi\rho(0)$ are kept fixed as
$N\!\to\!\infty$ this {\em Gaussian} random matrix partition function
equals the Leutwyler-Smilga chiral Lagrangian (\ref{ZLS}) provided one
makes the identification $\Sigma = 2\pi\rho(0)$. This exact
equivalence between the random matrix partition function and the QCD
effective Lagrangian in this regime, eq. (\ref{ZLS}), is crucial for the
understanding of why random matrix theory can be used to compute Dirac
operator spectra. It is therefore important to notice that also this
equivalence holds {\em universally}, independent of the choice of the
random matrix potential (up to the restrictions specified in refs.
\cite{ADMN,ADMN'}). We shall now prove this.


\section{Universality of the partition functions} 

There are at least two ways to demonstrate universality of the random matrix
partition function, and, subsequently, the identity
(up to an irrelevant $\mu_i$-independent constant):
\beq
{\cal Z}_{\nu}^{(N_{f})}(\mu_1,\ldots,\mu_{N_{f}}) ~=~ 
\tilde{\cal Z}_{\nu}^{(N_{f})}(\mu_1,\ldots,\mu_{N_{f}}) ~, 
\label{ZZ}
\eeq 
Here the l.h.s. is the finite-volume QCD partition function (\ref{ZLS}),
and the r.h.s. is the random matrix partition function (\ref{ZRM})
evaluated in the double-microscopic limit. The most direct way is to
use the orthogonal polynomial representation of the latter, $i.e$ define
monic orthogonal polynomials $P_n(\tila;m_1,\ldots,m_{N_{f}})$ so that
(expressed in terms of $\tila = \la^2$, which are more convenient
variables in the random matrix context):
\beq
\int_0^{\infty}d\tila ~\tila^{\nu}\prod_f(\tila+m_f^2)e^{-NV(\tila)}
P_k(\tila;m_1,\ldots,m_{N_{f}})P_{\ell}(\tila;m_1,\ldots,m_{N_{f}})
~=~ h_k(m_1,\ldots,m_{N_{f}})\delta_{k\ell} ~.\label{Pol}
\eeq
The random matrix partition function is directly related to the 
normalization constants $h_k$:
\beq
\tilde{\cal Z}_{\nu}^{(N_{f})} = N!\prod_{k=0}^{N-1}h_k(m_1,\ldots,
m_{N_{f}}) = N!h_0(m_1,\ldots,m_{N_{f}})^N
\prod_{k=1}^{N-1}r_k(m_1,\ldots,m_{N_{f}})~,
\label{Znorm}
\eeq
where $r_k \equiv h_k/h_{k-1}$. 
It is now possible to use the universality proof of ref. \cite{ADMN},
extended to the case of finite masses \cite{DN}, to prove the universal
relation (\ref{ZZ}) in the double-microscopic limit. This is most easily
done by taking the logarithm of (\ref{Znorm}), and turning the resulting
sum into an integral in the large-$N$ limit.

\noi
A much simpler way to prove universality of (\ref{ZZ}) is to make use of
an interesting relation between the orthogonal polynomials of eq. (\ref{Pol})
and the random matrix partition functions \cite{DAD}:
\beq
P_{N}(-\mu_{N_{f}+1}^2;\mu_1,\ldots,\mu_{N_{f}}) ~=~ C
(-1)^N(\mu_{N_{f}+1})^{-\nu}\
\frac{{\cal Z}^{(N_{f}+1)}_\nu(\mu_1,\ldots,\mu_{N_{f}+1})}
{{\cal Z}^{(N_{f})}_\nu(\mu_1,\ldots,\mu_{N_{f}})} ~. \label{polzch}
\eeq
Here $C$ is an irrelevant $\mu_i$-independent constant that just fixes
the overall normalization of the polynomials.
The l.h.s. of eq. (\ref{polzch})
was proven to have a universal double-microscopic limit in
refs. \cite{ADMN,DN} (and the solution is unambiguously defined also for 
negative eigenvalue entries). The universality proof now proceeds 
recursively, or by induction. For $N_f=0$ (the quenched case) universality
holds trivially. Universality of the random matrix partition
functions for $N_f=1$, and higher, then follows recursively, using
eq. (\ref{polzch}). Of course, this proves only universality of the 
result from the random matrix side, but not the identity (\ref{ZZ}). 
Fortunately, the identity has been established for the special case of 
Gaussian potentials in refs. \cite{SV,HV}. The simple proof given here
therefore extends this identity to the full universality class.

\section{Spectral correlators from partition functions}

The very useful connection between the orthogonal
polynomials and random matrix partition functions (\ref{polzch}) is 
actually only one in a series of such relations \cite{DAD,NDW}. Taken
separately, these relations provide very convenient and compact expressions
for all the relevant objects that enter in the random matrix computations.
But when used in connection with the identity (\ref{ZZ}) they provide
a much more intriguing series of relations -- relations that now only refer to
the finite-volume QCD partition functions.

\noi
Most important in this context is the corresponding partition function
representation of the kernel in random matrix theory:
\beq
K_N(\la,\la';m_1,\ldots,m_{N_{f}}) ~=~ 
e^{-\frac{N}{2}(V(\la^2)+V(\la'^2))}(\la\la')^{\nu+\frac{1}{2}}
\prod_{f}\sqrt{(\la^2+m_f^2)(\la'^2+m_f^2)}
\sum_{i=0}^{N-1} P_i(\la^2)P_i(\la'^2) ~.\label{kernel}
\eeq 
As is well known, from this kernel one can derive
all spectral correlation functions in the limit $N\to\infty$:
\beq
\rho(\tila_1,\ldots,\tila_n;m_1,\ldots,m_{N_{f}}) ~=~
\det_{a,b} K(\tila_a,\tila_b;m_1,\ldots,m_{N_{f}}) ~.
\label{correl}
\eeq

\noi
We can now make use of the following convenient representation of the kernel:
\begin{eqnarray}
&&K_N(z,z';m_1,\ldots,m_{N_{f}}) =
\frac{e^{-\frac{N}{2}(V(z^2)+V(z'^2))}(zz')^{\nu+\frac{1}{2}}
\prod_{f}^{N_{f}}\sqrt{(z^2+m_f^2)(z'^2+m_f^2)}}{
\tilde{\cal Z}_{\nu}^{(N_{f})}(m_1,\ldots,m_{N_{f}})} \cr
&&~\times \prod_f^{N_f}(m_f^{\nu})
\int_0^{\infty}\! \prod_{i=1}^{N-1}\!\!\left(\!d\lambda_i
\la_i^{\nu}(\la_i-z^2)(\la_i-z'^2)\!\!\prod_{f}^{N_{f}}(\lambda_i + m_f^2)
\mbox{e}^{-NV(\lambda_i)}\!\right)\!\left|{\det}_{ij}
\lambda_j^{i-1}\right|^2 ~.\label{Krep}
\end{eqnarray}
Except for the fact that the last eigenvalue integral runs up to $N-1$
only, the last factor is simply yet another partition function, now with
two additional quark species of imaginary mass! Thus,
up to corrections of order $1/N$, we have in the large-$N$ limit: 
\begin{eqnarray}
K^{(N_{f},\nu)}_N(z,z';m_1,\ldots,m_{N_{f}}) &=&
e^{-\frac{N}{2}(V(z^2)+V(z'^2))}(-1)^{\nu}\sqrt{zz'}
\prod_{f}^{N_{f}}\sqrt{(z^2+m_f^2)(z'^2+m_f^2)}\cr
&&\times ~\frac{
\tilde{\cal Z}_{\nu}^{(N_{f}+2)}(m_1,\ldots,m_{N_{f}},iz,iz')}{
\tilde{\cal Z}_{\nu}^{(N_{f})}(m_1,\ldots,m_{N_{f}})} ~,
\label{Krep1}
\end{eqnarray}

\noi
We are now ready to take the double-microscopic limit in which
$\zeta \equiv z N2\pi\rho(0)$ and $\mu_i \equiv m_i N2\pi\rho(0)$ are kept 
fixed as $N\!\to\!\infty$. In this limit the prefactor
$\exp[-(N/2)(V(z^2)+V(z'^2))]$ becomes replaced by unity. By
identifying $\Sigma = 2\pi\rho(0)$, and using the universal relation
(\ref{ZZ}) we finally arrive at the following master formula \cite{DAD}:
\beq
K_S^{(N_{f})}(\zeta,\zeta';\mu_1,\ldots,\mu_{N_{f}}) ~=~ C_{2}
\sqrt{\zeta\zeta'}\prod_{f}^{N_{f}}
\sqrt{(\zeta^2+\mu_f^2)(\zeta'^2+\mu_f^2)}~\frac{
{\cal Z}_{\nu}^{(N_{f}+2)}(\mu_1,\ldots,\mu_{N_{f}},i\zeta,i\zeta')}{
{\cal Z}_{\nu}^{(N_{f})}(\mu_1,\ldots,\mu_{N_{f}})} ~.\label{mf}
\eeq
{}From this one single formula all double-microscopic spectral correlators
can be computed directly from QCD chiral Lagrangians in the appropriate
scaling regime. In particular, for the spectral density itself we find
\beq
\rho_S(\zeta;\mu_1,\ldots,\mu_{N_{f}}) ~=~
C_2 |\zeta| \prod_{f}^{N_{f}}(\zeta^2+\mu_f^2)~\frac{
{\cal Z}_{\nu}^{(N_{f}+2)}(\mu_1,\ldots,\mu_{N_{f}},i\zeta,i\zeta)}{
{\cal Z}_{\nu}^{(N_{f})}(\mu_1,\ldots,\mu_{N_{f}})} ~.\label{spec}
\eeq
The overall proportionality factor $C_2$ can be determined by using the
matching condition 
\beq
\lim_{\zeta\to\infty} 
\rho_S(\zeta;\mu_1,\ldots,\mu_{N_{f}}) = 1/\pi ~, 
\eeq
which fixes $C_2 = (-1)^{\nu+[N_{f}/2]}$.

\noi
The higher $k$-point double-microscopic spectral correlation functions 
are conveniently evaluated using the double-microscopic limit
of the general relation (\ref{correl}). Curiously, it is also possible
to relate these higher $k$-point functions to finite-volume QCD partition
functions with $2k$ additional quark species \cite{DAD}:
\beqn
\rho_S(\zeta_1,\ldots,\zeta_k;\mu_1,\ldots,\mu_{N_f})
&=& C^{(k)}
\prod_i^k\left(\zeta_i\prod_f^{N_f}(\zeta_i^2+\mu_f^2)\right)
\prod_{j<l}^k|\zeta_j^2-\zeta_l^2|^2 \nonumber\\
&&\times\
\frac{{\cal Z}_{\nu}^{(N_{f}+2k)}
(\mu_1,\ldots,\mu_{N_f};\{i\zeta_1\},\ldots, \{i\zeta_k\})}
{{\cal Z}_{\nu}^{(N_{f})}(\mu_1,\ldots,\mu_{N_f})} ,
\label{corrft}
\eeqn
Each additional imaginary quark mass $i\zeta_j$ is thus doubly degenerate.
The overall proportionality constant $C^{(k)}$ can again be fixed by a 
matching condition. For $k=1$ the relation (\ref{corrft}) simply coincides
with the previous expression for the double-microscopic spectral density.
But already for $k=2$ (and all higher values of $k$) the expressions are
completely different, relating as they do the spectral correlators to
finite-volume QCD partition functions with different numbers of flavors.
It is quite amazing that the finite-volume QCD partition function 
(\ref{ZLS}) has all this structure, which takes on such a simple form
in random matrix language, encoded in it. In fact, by combining
eqs. (\ref{correl}) and (\ref{corrft}) one obtains an infinite sequence
of consistency conditions for QCD partitions. The relations 
become particularly transparent if we first take the additional fermion
masses to physical values by replacing $\zeta_j\to -i\zeta_j$ (inspection
of the explicit solution of ref. \cite{JSV} shows immediately that
this can be done unambiguously). We then
find the following infinite sequence of consistency conditions \cite{DAD}:
\begin{eqnarray}
&&\det_{1\leq a,b\leq k}\left[\sqrt{\zeta_a\zeta_b}\prod_{f=1}^{N_{f}}
\sqrt{(\mu_f^2-\zeta_a^2)(\mu_f^2-\zeta_b^2)}
{\cal Z}_{\nu}^{(N_{f}+2)}(\mu_1,\ldots,\mu_{N_{f}},\zeta_a,\zeta_b)\right]
= \cr && \tilde{C}^{(k)}
\prod_i^k\left( \zeta_i\prod_{f=1}^{N_{f}}
(\mu_f^2-\zeta_i^2)\right)
\prod_{j<l}^k|\zeta_j^2-\zeta_l^2|^2 ~
\frac{{\cal Z}_{\nu}^{(N_{f}+2k)}
(\mu_1,\ldots,\mu_{N_f},\{\zeta_1\},\ldots,\{\zeta_k\})}
{{\cal Z}_{\nu}^{(N_{f})}(\mu_1,\ldots,\mu_{N_{f}})^{1-k}} ~,
\label{cons}
\end{eqnarray}
where $\tilde{C}^{(k)}$ is some overall $\mu_i$-independent normalization
constant. Precisely these relations encode in the finite-volume QCD
partition function the fact that in the random matrix picture the kernel 
(\ref{kernel}) generates all spectral correlation functions through the
relation (\ref{correl}).

\section{Direct computations from chiral Lagrangians}

We have learned that the {\em massless} 
spectral sum rules \cite{LS} do not provide the proper 
starting point for computing the microscopic spectral density. It is the
{\em double-microscopic} limit \cite{SV,JNZ,DN,WGW} that is needed.
This was in fact clear already from the first demonstration of the
equivalence of the random matrix theory partition function and the
finite-volume QCD partition function \cite{SV,HV}. Instead of the 
massless spectral sum rules, one should focus on the 
``massive spectral sum rules''
\cite{SV,D} because these contain the analytical structure that allows
one to unravel the spectral correlators from the finite-volume 
partition function. This fact becomes very clear when one considers the 
most simple example, the massive spectral sum rule corresponding to
{\em quenched} QCD. Defining
\beq
G(\mu) ~\equiv~ 2\mu\int_0^{\infty}d\lambda \frac{\rho_S(\lambda)}{
\la^2+\mu^2} ~,\label{G}
\eeq
this can be written as a Stieltjes transform:
\beq
\int_0^{\infty}dt \frac{\rho_S(t)/\sqrt{t}}{t+y} ~=~ G(\sqrt{y}) 
~\equiv~ F(y) ~.
\eeq
The inverse of this is given by the discontinuity:
\beq
\frac{\rho_S(\sqrt{t})}{\sqrt{t}} = \frac{1}{2\pi i}\lim_{\epsilon\to 0}
\left[F(-t-i\epsilon) - F(-t+i\epsilon)\right] ~. \label{disc}
\eeq
So if one could compute the l.h.s. of (\ref{G}) directly from a 
finite-volume partition function, one would have achieved a derivation
of the spectral density without having at any intermediate 
stage to go through the
random matrix theory framework at all. The trouble is that
the massive spectral sum rule (\ref{G}) refers to a ``quenched quark'' of
(rescaled) mass $\mu$. If it were a {\em dynamical} quark, one could
compute the function $G(\mu)$ straightforwardly from \cite{D}
\beq
G(\mu) ~=~ \frac{\partial}{\partial\mu} \ln Z_{\nu}(\mu) - \frac{\nu}{\mu} ~,
\label{G1}
\eeq
where the last term subtracts the contribution from the zero modes.
The needed trick, recently discovered by Osborn, Toublan and Verbaarschot
\cite{OTV}, is to compute the r.h.s. of eq. (\ref{G1}) in a finite-volume
field theory that contains yet another ``quark'', now of opposite
statistics and of initially different mass (so that, after taking the
degenerate mass limit, the two determinants cancel in the partition function
itself). The result is (for $\nu=0$) \cite{OTV}:
\beq
G(\mu) = \Sigma\mu[I_0(\mu)K_0(\mu)+ I_1(\mu)K_1(\mu)]
\eeq
a result that was first derived the other way around, from the random
matrix theory result, by Verbaarschot \cite{V'}. Substituting this into
eq. (\ref{disc}), one finds, straight from the finite-volume partition
function,
\beq
\rho_S(\la) = \frac{1}{2}|\la|\left[J_0(\la)^2 + J_1(\la^2)\right]
\eeq
which of course agrees with the result obtrained from random matrix theory.
Not only could those authors compute the function $G(\mu)$ this
way, they also managed to rewrite the general expression (\ref{disc})
in precisely the form (\ref{spec}) using the technique of partially
quenched chiral perturbation theory \cite{BG}, here based on the
super Lie group $U(N_f+1|1)$. Their result naturally generalizes
to higher $k$-point spectral correlation functions, now given in the
form (\ref{corrft}). The relevant super Lie group will here be
$U(N_f+k|k)$.

\section{Finite-volume corrections}

We have seen that the microscopic spectral correlators have a natural
interpretation as finite-size scaling functions.
This makes them ideally suited for lattice gauge theory studies,
and in fact there have now been Monte Carlo tests of the spectral
densities of QCD in both (3+1) dimensions \cite{MC4} and (2+1) dimensions
\cite{MC3}. (There have also been interesting studies of the applicability
of random matrix techniques beyond the microscopic limit, in the ``bulk''
\cite{bulk}). One obvious question in that connection concerns 
finite-size corrections. In actual computations the volume $V$
is often far from being asymptotically large, and one could ask whether
it is also possible to analytically calculate subleading corrections.
For example, for the double-microscopic spectral density itself one could
envisage an expansion of the kind
\beq
\rho_S^{(V)}(\zeta;\mu_1,\ldots,\mu_{N_{f}}) =
\rho_S^{(\infty)}(\zeta;\mu_1,\ldots,\mu_{N_{f}})
\left[1 + \frac{A}{V}f(\zeta;\mu_1,\ldots,\mu_{N_{f}}) + \ldots\right] ~,
\eeq
with $A$ some dimensionful constant, and $f$ a correction-to-scaling
function. One could hope that such corrections could be computed analytically
using the random matrix theory formulation. In fact, if we go back to the
derivation of eq. (\ref{Krep1}) from (\ref{Krep}), we could try to keep
the subleading corrections that come from ignoring the difference between
$N$ and $N-1$ in (\ref{Krep}). However, on top of these we can also get
subleading contributions from the potential $V(\la^2)$, even in the
microscopic limit. We therefore conclude that such subleading $1/N$
corrections in the random matrix picture will be {\em non-universal}, and
hence cannot be expected to be related to the Dirac eigenvalue spectrum.
This is completely in accord with the field theory picture, in which
subleading terms in $1/V$ will involve non-static modes of the
pseudo-Goldstone bosons. The kinetic term 
$\frac{1}{2}f_{\pi}^2 Tr[\partial_{\mu}U\partial_{\mu}U]$ in the effective
Lagrangian can therefore not be neglected. A new dimensionful
scale (namely $f_{\pi}$) has entered, and the Dirac spectrum will cease to
be a scaling function related to just $V$ and $\Sigma$. Of course, one
could try to systematically analyze $1/V$ corrections in this very precise
framework of the effective Lagrangian. The most natural starting point
will unfortunately not be the usual expansion around the kinetic term, 
but rather a low-momentum expansion around the mass term Tr$[{\cal M}U]$.   

\section{Flavor dependence}
 
We finally address the question of the flavor dependence of all these
results. The number of flavors enters in a very simple way in both the
field theory and random matrix picture. In the former it determines the
coset integration for the effective Lagrangian, while in the latter it
enters only through the strength of the determinant in the
expression (\ref{ZRM}). Both lead to an extremely mild dependence on
the number of flavors. This is because we throughout normalize the
chiral condensate to one flavor-independent number $\Sigma$, -- a convenient
normalization because it puts the different theories with different
flavor content on the same common scale. Clearly the whole framework
collapses as the number $N_f$ exceeds the value $N_f^*$ above which
QCD no longer supports spontaneous chiral symmetry. If we allow ourselves
to treat this upper number of flavors $N_f^*$ as a free and tunable parameter,
then a normalizable condensate $\Sigma$ will simply cease to exist 
precisely at $N_f=N_f^*$. In the random matrix theory context this may
correspond to hitting the boundary where $\rho(0) \to 0$ \cite{ADMN'}.
Beyond this point all results discussed here will no longer be valid.
Just at the point where the condensate disappears one can define critical
exponents that count the rate at which $\rho(\la)$ vanishes as $\la\to 0$
\cite{JNPZ}; however they will here not correspond to a physical
phase transition (appearing instead as one tunes $N_f$ continuously
to $N_f^*$). Beyond that point the spectral density of the Dirac operator
will develop a gap around $\la=0$, and there is no obvious way to compute
it from the finite-volume partition function (mesons with quantum numbers
of the pseudo-Goldstone bosons
will still be the lightest excitations \cite{W}, but there will be no
analogue of the chiral Lagrangian).

\noi
There is however a limit of large $N_f$ in which the results here will
continue to be valid. This is what is called the ``topological $1/N$ 
expansion'', where $N_f\to\infty$ and $N_c\to\infty$, with the
ratio $\eta \equiv N_f/N_c$ kept fixed. The pertinent random matrix
theory ensembles (and chiral Lagrangians) continue to be the same, and
if $\eta$ is chosen small enough (as in QCD), the theory will still
undergo spontaneous chiral symmetry breaking. However, because $N_f\to
\infty$, there is now non-trivial ``dynamics'' even in the very simple
chiral Lagrangian (\ref{ZLS}). In fact, when all masses are chosen equal
the theory becomes identical to large-$N_c$ lattice QCD in (1+1) dimensions 
will so-called Wilson action. It is known that this theory undergoes a 
3rd order phase transition \cite{GW}, and the QCD partition function 
(\ref{ZLS}) therefore undergoes precisely such a phase transition in
the above limit. Because of the connection between partition functions
and the microscopic spectral correlators, such a phase transition is
expected to surface also in the Dirac spectrum, once the above limit is
taken. Because a new limit ($N_f\to\infty$ and $N_c\to\infty$, with 
$\eta \equiv N_f/N_c$ fixed) is taken on top of the usual double-microscopic
limit, we need to redefine our scaling variables. Let $\kappa
\equiv N_f/\mu$, and keep this variable fixed as $N_f\to\infty$. If we
define the free energy by ${\cal F}\equiv -[\ln{\cal Z}^{(N_{f})}]/N_f^2$,
then from the Gross-Witten analysis we know that
\begin{eqnarray}
{\cal F} &=& 1/(4\kappa^2) ~~, ~~ \kappa \geq 1 \cr
{\cal F} &=& 1/\kappa + (1/2)\ln(\kappa) - 3/4 ~~, ~~ 
\kappa \leq 1 ~.
\end{eqnarray}
Taking derivatives, one indeed verifies that only the 3rd derivative of
${\cal F}$ is discontinuous at $\kappa=1$. It is amusing that this 3rd
order phase transition of the effective QCD partition function has a
simple interpretation as occurring in a two-dimensional ``world volume''
taken to be the 2-dimensional 
space of the $N_f\times N_f$ unitary matrices $U$ \cite{DH}. 
One can plot the complex zeros of the partition function, and see that
they as expected precisely pinch the real $\kappa$-axis at $\kappa=1$.
From this one can very accurately compute the correlation length critical 
index $\nu$, which turns out to agree with an analytical result obtained
from the so-called double-scaling limit of models of 2-d quantum gravity
\cite{PS}. More importantly, the critical scaling corresponding to this
critical index sets in at very, very low values of $N_f$ -- actually already
at $N_f=2$ \cite{DH}. Traces of this analog of the Gross-Witten
phase transition therefore persist even in real (3+1)-dimensional QCD,
in the microscopic scaling regime.  

\noi
{\sc Acknowledgements:} I thank G. Akemann, U. Heller, U. Magnea, S. Nishigaki,
and T. Wettig for stimulating discussions in connection with our joint work.


\end{document}